\documentclass[twocolumn,showpacs,preprintnumbers,nofootinbib,prd,
superscriptaddress,10pt]{revtex4-1}

\usepackage{amsmath,amssymb}
\usepackage{amsfonts}
\usepackage[normalem]{ulem}
\usepackage{textcomp}
\usepackage{hyperref}
\usepackage{enumitem}
\usepackage{bm}
\usepackage{afterpage}
\usepackage{graphicx}
\usepackage{psfrag}
\usepackage{mathtools}
\usepackage{tensor}
\usepackage{layouts}
\usepackage{DejaVuSans}
\usepackage{epstopdf}
\usepackage[usenames,dvipsnames]{xcolor}
\usepackage[utf8]{inputenc}
\usepackage{multirow}
\usepackage{rotating}
\usepackage{tabularx}
\usepackage{ragged2e}
\usepackage{blindtext}
\usepackage{graphicx}
\usepackage{siunitx}
\newcommand{\R}{\mathbb{R}}

\DeclareMathOperator{\sign}{sign}
\renewcommand{\d}[1]{\ensuremath{\operatorname{d}\!{#1}}}

\begin{document}

\begin{abstract}
We apply machine learning methods to build a time-domain model for
gravitational waveforms from binary black hole mergers, called \texttt{mlgw}.
The dimensionality of the problem is handled by representing the 
waveform's amplitude and phase using a principal component analysis.
We train \texttt{mlgw} on about $\mathcal{O}(10^3)$ \texttt{TEOBResumS} and \texttt{SEOBNRv4} effective-one-body
waveforms with mass ratios $q\in[1,20]$ and aligned dimensionless
spins $s\in[-0.80,0.95]$. The resulting models are faithful to the
training sets at the ${\sim}10^{-3}$ level (averaged on the parameter space).
The speed up for a single waveform generation is  a factor 10 to 50 (depending 
on the binary mass and initial frequency) for \texttt{TEOBResumS} and
approximately an order of magnitude more for \texttt{SEOBNRv4}.
Furthermore, \texttt{mlgw} provides a closed form expression for the waveform 
and its gradient with respect to the orbital parameters; such an information might 
be useful for future improvements in GW data analysis. 
As demonstration of the capabilities of \texttt{mlgw} to perform a
full parameter estimation, we re-analyze the public data from the first
GW transient catalog (GWTC-1). We find broadly consistent results with previous 
analyses at a fraction of the cost, although the analysis with spin aligned waveforms 
gives systematic larger values of the effective spins with respect to previous analyses 
with precessing waveforms. Since the generation time does not depend 
on the length of the signal, our model is particularly suitable for the analysis 
of the long signals that are expected to be detected by third-generation detectors. 
Future applications include the analysis of waveform systematics and
model selection in parameter estimation.

\end{abstract}
	
	\title{Machine Learning Gravitational Waves
            from Binary Black Hole Mergers}
	\author{Stefano \surname{Schmidt}}
		\email{stefanoschmidt1995@gmail.com}
        \affiliation{Dipartimento di Fisica  Universit\`a di Pisa, and INFN Sezione di Pisa, Pisa I-56127,Italy}        
	\author{Matteo \surname{Breschi}}
        \affiliation{Theoretisch-Physikalisches Institut, Friedrich-Schiller-Universit{\"a}t Jena, 07743, Jena, Germany}
        \author{Rossella \surname{Gamba}}
        \affiliation{Theoretisch-Physikalisches Institut, Friedrich-Schiller-Universit{\"a}t Jena, 07743, Jena, Germany}
        \author{Giulia \surname{Pagano}}
        \affiliation{Dipartimento di Fisica  Universit\`a di Pisa, and INFN Sezione di Pisa, Pisa I-56127,Italy} 
        \author{Piero \surname{Rettegno}}
        \affiliation{Dipartimento di Fisica, Universit\`a di Torino, via P. Giuria 1, 10125 Torino, Italy}
        \affiliation{INFN Sezione di Torino, Via P. Giuria 1, 10125 Torino, Italy}
        \author{Gunnar \surname{Riemenschneider}}
        \affiliation{Dipartimento di Fisica, Universit\`a di Torino, via P. Giuria 1, 10125 Torino, Italy}
        \affiliation{INFN Sezione di Torino, Via P. Giuria 1, 10125 Torino, Italy}
        \author{Sebastiano \surname{Bernuzzi}}
        \affiliation{Theoretisch-Physikalisches Institut, Friedrich-Schiller-Universit{\"a}t Jena, 07743, Jena, Germany}
        \author{Alessandro \surname{Nagar}}
        \affiliation{INFN Sezione di Torino, Via P. Giuria 1, 10125 Torino, Italy}
        \affiliation{IHES, 91440, Bures-sur-Yvette, France}
        \author{Walter \surname{Del Pozzo}}
        \affiliation{Dipartimento di Fisica  Universit\`a di Pisa, and INFN Sezione di Pisa, Pisa I-56127,Italy}        
	
	\maketitle

\section{Introduction}
The detection of Gravitational Waves (GW) from compact binary coalescence (CBC) has been possible thanks to the joint effort of a number of different fields of expertise, all joining together to achieve the sophisticated detection process. GW data analysis concerns the detection of a GW signal hidden in the raw detector output (\textit{matched filtering}) and subsequently the inference of its physical properties (\textit{parameter estimation}). In order to accomplish its goal, GW data analysis relies on the availability of waveform (WF) templates to compare with the detector output.
To accurately explore the posterior distribution for the parameters defining a CBC, state-of-the-art parameter estimation (PE) algorithms~\cite{Aasi:2013jjl}~\cite{Veitch2014wba} can require the generation of as many as $10^7$ waveform templates. It is therefore paramount for the waveform generation to be as fast as possible. At the same time, because of the extreme sensitivity to phase differences in the likelihood  function, 
the templates must retain a high degree of accuracy to avoid biases in the posterior exploration.

Many efforts have been devoted to numerically solve Einstein equations for two coalescing objects and 
to predict the gravitational radiation emitted~\cite{Boyle:2019kee,Healy:2019jyf,Healy:2020vre}.
As solving the full equations is still extremely computationally challenging, the LIGO-Virgo Collaboration 
relies on approximate analytical models. 
These can be broadly categorized in  three families; (i) effective-one-body (EOB)~\cite{Buonanno:2000ef} 
waveform models~\cite{Damour:2009kr,Nagar:2020pcj,Chiaramello:2020ehz,Ossokine:2020kjp}; (ii) phenomenological 
models~\cite{Khan:2015jqa,Pratten:2020ceb,Estelles:2020osj}; (iii) NR surrogates~\cite{Varma:2018mmi,Williams:2019vub,Varma:2019csw,Rifat:2019ltp,Khan:2020fso}.

EOB models are the waveform approximants that incorporate the largest amount of analytical information.
They compute the GW signal by solving Hamilton's equations  and accurately predict the 
GW signal from early inspiral phase up to the final ringdown. The underlying relative dynamics
is informed by (or calibrated to) NR simulations via a few parameters that take into account 
in an effective way yet uncalculated high-order corrections to the orbital or spin-orbit sector 
of the Hamiltonian. Similarly, the merger and ringdown parts of the waveform are also informed by NR 
simulations\footnote{One should remember that this step is a priori not necessary in the theoretical 
construction of the model~\cite{Buonanno:2000ef,Damour:2007xr,Damour:2009wj}.}.
Because of the numerical integrations involved to solve Hamilton's equations, they tend to 
be accurate, but sometimes slow to generate, see however~\cite{Nagar:2018gnk} for a 
more efficient approach to obtain the dynamics.

The phenomenological waveforms are based on the post-Newtonian formalism and then calibrated 
to  EOB waveforms and numerical relativity. They tend to be faster than EOB models, but not as accurate.

Many efforts have been devoted to the task of speeding up the generation of GW signals from 
EOB families. For example, one lead to the development of \textit{surrogate models}. Surrogate 
models are constructed starting from some decomposition in the waveform space followed by efficient 
interpolation to avoid any  numerical integration~\cite{Purrer:2015tud,Bohe:2016gbl,Purrer:2017str,Lackey:2018zvw,Cotesta:2020qhw}. 
Being fast to generate, they are routinely employed in GW data analysis. 

A Machine Learning model is a promising alternative to accelerate the
waveforms generation of state-of-the-art models.
Machine Learning (ML) is a branch of statistics that is devoted to reproduce patterns read from data. 
A ML algorithm needs very little human input and, by automatically solving an optimization problem, 
it is able to choose the best performing element among a large class (i.e. the \textit{model}) of parametric solutions. 
This is the so-called training procedure. 
The ML flexibility in modeling data and reproducing trends is appealing: with a proper model choice 
and with an appropriate training procedure, we can hope to have a reliable, fast to execute generator 
of GW waveforms, while retaining the accuracy necessary for robust parameter estimation.
ML procedures have already been successfully exploited for speeding up the WF generation~\cite{Khan:2020fso, Chua_2019}
and for signal detection and/or parameter estimation~\cite{Gabbard:2017lja,George:2017pmj,George:2017vlv,Rebei:2018lzh,Chatterjee:2019gqr,Wong:2020wvd, Khan:2020foe}. A comparative study of different regression methods for the task of generating WFs is performed in \cite{Setyawati:2019xzw}.

In this work, we explore the application of ML to WF generation and we build a ML model, called \texttt{mlgw}, that allows to reproduce 
waveforms from any EOB model for BBH coalescence. We demonstrate that such ML-based model can generate 
GW signals significantly faster than the original model, 
matching the performances of a Reduced Order Modelling 
(ROM)~\cite{Purrer:2015tud,Bohe:2016gbl,Purrer:2017str, Field:2013cfa}.  
At the same time it shows a good agreement with the train model, at the $10^{-3}$ level when 
averaged on the whole parameter space. For simplicity, we focus here only on the 
dominant $\ell=m=2$ quadrupole waveform and we only consider the aligned spin case.

EOB models are the ideal candidate for training our model: they are moderately fast to run 
and, although they are {\it checked} (and NR-informed) only on the limited part of the 
parameter space covered by NR simulations,  they are also typically able to robustly 
generate waveforms for a large set of parameters outside the NR-information 
domain (e.g. large mass ratios and spins).
This is the case of the most recent avatar of the spin-aligned \texttt{TEOBResumS}~\cite{Nagar:2018zoe} 
model, that incorporates subdominant waveform modes, \texttt{TEOBResumS\_SM}~\cite{Nagar:2019wds,Nagar:2020pcj}
\footnote{This model used several hundreds of the available SXS simulations to inform a highly 
accurate description of the postmerger-ringdown phase~\cite{Damour:2014yha}, but 
only around 40 datasets to improve the behavior of the analytical EOB dynamics.}.
\texttt{TEOBResumS\_SM} is NR-faithful over the largest set of spin-aligned NR waveforms 
available today (595 datasets)~\cite{Nagar:2020pcj}, with maximum EOB/NR unfaithfulness 
always below $0.5\%$. Here, to reduce the generation time of the training set, and to be more 
conservative, we slightly downgrade the performance of \texttt{TEOBResumS} considering 
its version {\it without} iteration~\footnote{This slightly worsens the EOB/NR performance that in
any case remains below $1\%$ except for $\sim 40$ outliers that still do not exceed the $3\%$ 
threshold and are mostly below the $2\%$ level~\cite{Riemenschneider:2020}.} on 
the next-to-quasi-circular parameters~\cite{Nagar:2020pcj}.
The other EOB-based model in 
use is the~\texttt{SEOBNRv4}~\cite{Taracchini:2013rva,Bohe:2016gbl} one, largely employed 
by the LIGO-Virgo collaboration. This model was calibrated to NR so as to get  
maximum EOB/NR unfaithfulness at most of $1\%$. However, differently from 
TEOB~\cite{Nagar:2018zoe} this model in its native implementation is computationally 
too slow for parameter estimation purposes and it is absolutely necessary 
to use its ROM version, ~\texttt{SEOBNRv4\_ROM}~\cite{Bohe:2016gbl}.

We use both \texttt{TEOBResumS\_SM} (in the following simply \texttt{TEOBResumS}) 
and \texttt{SEOBNRv4} to train different two different versions of \texttt{mlgw}, respectively
\texttt{mlgw-TEOBResumS} and \texttt{mlgw-SEOBNRv4}.
As a relevant physical application, we use both 
\texttt{mlgw-TEOBResumS} and \texttt{mlgw-SEOBNRv4}.
to provide two new, and independent, analyses of the 10 BBHs 
coalescence events collected in the O1/O2 LIGO-Virgo observing 
runs~\cite{LIGOScientific:2018mvr}. The outcome of the analyses 
using the two models are largely consistent among themselves.
In addition,  the so-obtained physical properties of the 10 BBHs
systems are compatible with previous analyses published in
Ref.~\cite{LIGOScientific:2018mvr}, though obtained using
different, notably spin-precessing, waveform models.

On top of the specific application discussed here, our ML waveform model could also be used directly 
to speed up GW data analysis.
Furthermore, since the time required to generate a WF does not depend
on the signal time length but only only on the number of grid points which the WF is evaluated at,
the applicability of our approach goes far beyond the LIGO/Virgo physics scenario. In particular,
we think about the forthcoming Einstein Telescope, that will be sensitive to very long stellar-mass 
inspirals from 5Hz, or to extreme mass ratio inspirals as LISA sources. In this context the problem
of WF fast generation will be more pressing and our approach, provided a suitable waveform model 
for training,  might be essential for detection and parameter estimation. 

The paper is organized as follows. In Sec.~\ref{sec:setup} we briefly set the 
notation and the core of the ML problem we solve; in Sec.~\ref{sec:model} we describe 
our model in details. Section ~\ref{sec:performance} is devoted to validate the model 
and to assess its accuracy and speed of execution; Sec.~\ref{sec:GWTC1} holds 
our analysis of the GWTC-1 transient catalog, while  Sec.~\ref{sec:end} collects 
some final remarks and future prospects of our work.

\section{Conventions setup}
\label{sec:setup}
A binary black hole system is parametrized by a vector $ \boldsymbol{\vartheta} = (m_1,m_2, \mathbf{s}_1,\mathbf{s}_2) $, where $m_i$ are the 
BHs masses and $\mathbf{s}_i \equiv \mathbf{S}_i/m_i^2 \leq 1$ are the 
\textit{dimensionless} spin. We call them the \textit{orbital parameters}.
We use the convention $m_1\geq m_2$ and we denote the total mass as $M\equiv m_1+m_2$
and the mass ratio as $q\equiv m_1/m_2$.
In what follows, we will focus on the case in which spins $\mathbf{s}_1$ and $\mathbf{s}_2$ are \textit{aligned} with the orbital angular momentum. 
Let $d_L$ be the luminosity distance and $\iota$ and $\varphi_0$ the polar angle (inclination) and the azimuthal angle of the orbital plane.
A GW is parametrized as~\cite[Eq.~II.6]{ajith2011data}:
\begin{align} \label{eq:h_parametrization}
	&h(t; d_L,\iota,\varphi_0, \boldsymbol{\vartheta}) = h_+ + i h_\times \nonumber \\
		&\qquad= \frac{G}{c^2} \frac{M}{d_L}\sum_{\ell = 2}^{\infty} \sum_{m = -\ell}^{\ell} \tensor[_{-2}]{Y}{_{\ell m}}(\iota, \varphi_0) H_{\ell m}(t/M; \boldsymbol{\tilde{\vartheta}})
\end{align}
where $\tensor[_{-2}]{Y}{_{\ell m}}(\iota, \varphi_0)$ are the spin-2 spherical harmonics. 
Once written as a function of the dimensionless time $t/M$, the quantities $H_{\ell m}$ 
depends only on the variables $\tilde{\boldsymbol{\vartheta}} = (q, s_1, s_2)$, and we
are considering here only $\ell=|m|=2$.
Since the dependence on the two angles, on the distance and total mass is known, 
for convenience we fix their value to ${\iota = \varphi_0 = 0}$, ${d_L = \SI{1}{Mpc}}$ 
and $M = 20M_\odot$ so to only work with waveforms ${h_{\rm FIT}(t; \boldsymbol{\vartheta}) = h(t; d_L = \SI{1}{Mpc}, \iota = \varphi_0 = 0, 
M = 20M_\odot, \boldsymbol{\tilde{\vartheta}}) }$:
\begin{align}  
	h_{\rm FIT}(t; \boldsymbol{\vartheta}) &\equiv 9.6 \times 10^{-19} \; \tensor[_{-2}]{Y}{_{2 2}}(0,0) H_{22}(t/\SI{20}{M_\odot}; \boldsymbol{\tilde{\vartheta}})  \nonumber \\
	&= 6 \times 10^{-19} \; H_{22}(t/(20M_\odot); \boldsymbol{\tilde{\vartheta}}) \label{eq:h_std} \; .
\end{align}
Finally, we express  $h_{\rm FIT}$ in terms of its amplitude and phase
\footnote{Note we adopt a nonstandard sign convention for the phase}:
\begin{equation} \label{eq:h_ML}
	h_{\rm FIT}(t; \boldsymbol{\vartheta}) = A(t; \boldsymbol{\tilde{\vartheta}}) e^{i \phi(t; \boldsymbol{\tilde{\vartheta}})} \; . 
\end{equation}
We may also write $f_{\boldsymbol{\tilde{\vartheta}}}(t)$ to denote a 
function $f(t;\boldsymbol{\tilde{\vartheta}})$ of time with parametric dependence on $\boldsymbol{\vartheta}$.
In what follows, $f$ stands as a placeholder for any of the functions $A_{\tilde{\boldsymbol{\vartheta}}}(t)$ 
and ${\phi}_{\tilde{\boldsymbol{\vartheta}}}(t)$.
With this definition, the full waveform can be expressed as:
\begin{align} 
	h&(t, d_L,\iota,\varphi_0; \boldsymbol{\vartheta}) = \frac{M}{\SI{20}{M_\odot}} \frac{\SI{1}{Mpc}}{d_L} \times  \nonumber \\
		&\times \Bigg\{ \frac{1+\cos^2\iota}{2} A_{\boldsymbol{\vartheta}}(t_M)  \cos[\phi_{\boldsymbol{\vartheta}}(t_M)+2\varphi_0]  \nonumber \\
		&+ i \cos\iota A_{\boldsymbol{\vartheta}}(t_M)
 \sin[\phi_{\boldsymbol{\vartheta}}(t_M)+2\varphi_0] \Bigg\}
\label{eq:h_parametrization_simple}
\end{align}
where $t_M = t \; \frac{M}{\SI{20}{M_\odot}}$.
Note that in the equation above, we split the real and the imaginary part of $h$ and we 
used the relation ${\tensor[_{-2}]{Y}{_{2 \pm 2}}(\iota, \varphi_0) = \sqrt{\frac{5}{64\pi}} \; (1 \pm \cos \iota)^2 e^{\pm i 2 \varphi_0}}$.
As a constant translation of $\phi_{\boldsymbol{\vartheta}}$ can be absorbed in the definition of $\varphi_0$ and does not affect the physics, we choose the convention that $\phi_{\boldsymbol{\vartheta}} = 0$ when the amplitude $A_{\boldsymbol{\vartheta}}$ 
has a maximum.

\section{\lowercase{\texttt{mlgw}}}
\label{sec:model}
The goal of the present work is to provide an accurate Machine Learning model which outputs the functions $A(t;\boldsymbol{\tilde{\vartheta}})$ and $\phi(t;\boldsymbol{\tilde{\vartheta}})$ (Eq.~\eqref{eq:h_std}~and~\eqref{eq:h_ML}), as generated by the state-of-the-art time domain WF models.
More formally, we seek a ML model that reliably reproduces the following map:
\begin{align}
	(q, s_1, s_2) &\longmapsto A_{(q, s_1, s_2)}(t) \label{eq:objective_amp}\\
	(q, s_1, s_2) &\longmapsto \phi_{(q, s_1, s_2)}(t) . \label{eq:objective_ph}
\end{align}

In the context of ML, our task reduces to performing two regressions from $\tilde{\boldsymbol{\vartheta}}$ to the amplitude and phase of the WF. 
A {\it regression} is a statistical method to infer the relationship between a set of ``independent variables" and a set of ``dependent variables". A {\it model} consists in a functional form for such relation, usually with many free parameters to be specified. By looking at the data, one should be able to make a proper choice for their value.

In order to be able to perform each regression, several steps are required.

\begin{enumerate}[label=(\Alph*)]
	\item \textit{Setting a time grid}. Each WF must be represented on a discrete time grid, which allows for efficient and reliable reconstruction on an arbitrary, user-given, grid. After this operation, the functions $A(t)$ and $\phi(t)$ are represented as vectors\footnote{
In ML jargon, this procedure is called preprocessing and aims to create a standard representation for all the data available (in our case the WFs).
}.
	\item \textit{Creating a dataset of WFs}. A large number of WFs must be generated on the the chosen time grid for a different number of orbital parameter $(q,s_1,s_2)$. This will form the training set for the model.
	\item \textit{Reducing the dimensionality of a WF}. In order to make the regression feasible, we build a low dimensional representation of the WF. This operation must be invertible: once a low dimensional representation is given, one should be able to reconstruct the higher dimensional WF.
	\item \textit{Learning a regression}. We train a model to perform the regression from $(q, s_1, s_2)$ to the low dimensional representation of the WF.
\end{enumerate}
We discuss these points in detail in what follows.

%
\subsection{The time grid}
Each function $f(t)$ to fit (i.e. amplitude and phase) must be represented by its values $\mathbf{f} \in \R^D$ on a discrete grid of $D$ points $\mathbf{t} \in \R^D$.
It is convenient to work in a grid of (dimensionless) reduced time $\boldsymbol{\tau} \equiv \mathbf{t}/M$.
The time grid is chosen with the convention that at $\tau=0$ the function $A(t;\boldsymbol{\vartheta})$ (i.e. the amplitude of the ${22}$ mode) has a peak.
Once a time grid is set, the vector $\mathbf{f}$ is defined as follows:
\begin{equation}
	\mathbf{f}({\tilde{\boldsymbol{\vartheta}}})_i = f_{{\tilde{\boldsymbol{\vartheta}}}}(\boldsymbol{\tau}_i) \;\;\;\;\;\; i = 1, \ldots D
\end{equation}

The value of $f$ at an arbitrary time must be found by interpolation and to make the interpolation effective,
we introduce a grid adapted to the function's variation.
Clearly an equally spaced grid over times is not the best choice since the amplitude has a very narrow peak at $\tau=0$.
A good solution is to build the $\tau$ grid ${\boldsymbol{\tau}}$ as:
\begin{equation} \label{eq:fig01.tau_grid}
	{\boldsymbol{\tau}}_i = \sign{\boldsymbol{\tilde{\tau}}_i} \times (|\boldsymbol{\tilde{\tau}}_i|)^{\frac{1}{\alpha}} \;\;\;\;\;\; i = 1, \ldots D
\end{equation}
where $\boldsymbol{\tilde{\tau}}_i$ are D equally spaced points in the range of interest and we call $\alpha$ \textit{distortion parameter}.
This choice ensures that more points are accumulated around the peak of amplitude.
As the phase has a rather regular behavior, it is not important to tune the time grid on it. For this reason, a single grid for amplitude and phase, tuned on the amplitude, is used.

The length of the time grid determines the maximum length of the WFs that the model can generate.
Let us define $\tau_{\rm min} = -{\boldsymbol{\tau}}_0 > 0 $ the starting point of the grid; thus each WF starts 
at a time $\tau_{\rm min} M$ before the merger. Note that $\tau_{\rm min}$ is an important hyperparameter, 
set by the user, which strongly impacts on the model applicability.
The minimum frequency in the signal as a function of $M, q$ and $\tau_{\rm min}$ is given 
approximately~\footnote{The expression is approximate because it is obtained within a Newtonian framework and does not consider spin effects.
Nevertheless, it gives an useful estimation of the range of the applicability of the model.}
by:
\begin{align}\label{eq:f_min}
	f_{\rm min} = \SI{151}{Hz}  \left( \frac{(1+q)^2}{q} \right)^{\frac{3}{8}}  \left( \frac{M_\odot}{M} \right)  \left(\frac{\SI{1}{ \frac{s}{M_\odot}}}{\tau_{\rm min}} \right)^{\frac{3}{8}} .
\end{align}

\subsection{Dataset creation}
\label{sec:trainingset}
As in any ML method, we must create a dataset before training a model.
In our case, the dataset consist in a matrix ${X \in \mathbf{Mat}(N,3+2D)}$ of $N$ waveform, which has the following form:
\begin{equation} \label{eq:dataset}
	X_{i:} = [q,s_1,s_2, \boldsymbol{A}_{{\tilde{\boldsymbol{\vartheta}}}}^T, \boldsymbol{\phi}_{{\tilde{\boldsymbol{\vartheta}}}}^T]
\end{equation}
where $X_{i:}$ denotes the i-th row of the dataset matrix.

The dataset is filled with parameters ${\tilde{\boldsymbol{\vartheta}}}$ randomly drawn from an uniform distribution in the domain of interest $\mathcal{P}$: ${\tilde{\boldsymbol{\vartheta}}}_i \sim \textrm{Unif}(\mathcal{P})$.
As stressed above, any time domain EOB waveform model is suitable for such purpose. Indeed we employed 
successfully both\footnote{For completeness, we have also computed a {\tt mlgw} 
mode using \texttt{SEOBNRv2\_opt}~\cite{Devine:2016ovp}  a spin-aligned model that was optimized
with respect to the original \texttt{SEOBNRv2}~\cite{Taracchini:2013rva} 
so to improve its computational efficiency.} \texttt{TEOBResumS} and \texttt{SEOBNRv4}.
The output of the training model must be interpolated to the chosen time grid.

It is important to ensure that all waves have zero phase at a constant
time point $\bar{t}$: this is crucial to obtain a continuous
dependence of the phase components on the orbital parameters. As model
performances are not seen to depend on the choice of
$\bar{t}$, we arbitrarily set $\bar{t} = 0$.
The range $\mathcal{P}$ of masses and spins covered by the model, as 
well as the starting point of the grid $\tau_{\rm min}$, can be freely choose 
by the user, depending on their needs.

\subsection{Dimensionality reduction}
Once we are able to represent waveforms, a regressions ${\tilde{\boldsymbol{\vartheta}}} \longmapsto \boldsymbol{A}_{{\tilde{\boldsymbol{\vartheta}}}}, \boldsymbol{\phi}_{{\tilde{\boldsymbol{\vartheta}}}}$ is unfeasible, as the dimension of the target space is too large. Luckily, the elements of $\boldsymbol{A}, \boldsymbol{\phi}$ are strongly correlated with each other: the independent amount of information, required to fully reconstruct the wave, can be stored in a low dimensional vector.
A number of ML techniques to perform such a task are available. Among them, Principal Component Analysis (PCA) \cite[ch. 12]{murphy2012machine} was found to be particularly effective.

The basic idea behind PCA is to seek a \textit{linear relation} between high dimensional and low dimensional data: high dimensional data ($\in \R^D$) are projected onto a $K$ dimensional subspace, by means of an orthogonal projection.
A theorem~\cite[Sec. 12.2.1]{murphy2012machine} guarantees that, for zero mean data, the generators of subspace are the (orthonormal) first $K$ eigenvectors of the empirical covariance matrix $\Sigma \in \mathbf{Mat}(D,D)$. The eigenvectors are also called Principal Components (PCs) of the data.
Thus, the projection matrix $H\in \mathbf{Mat}(K,D)$ holds in each row the PCs and each high-dimensional point can be effectively expressed as a linear combination of the $K$ PCs~\footnote{For this reason, PCA can also be seen as a perturbative expansion of a high dimensional observation. A more reliable reconstruction can be achieved by adding more and more PCs, each of which is less important than its previous.}.

A PCA model is trained with the dataset Eq.~\eqref{eq:dataset}: it represents an (approximate) bijective map between the high dimensional WF $\mathbf{f} = \boldsymbol{A}_{\tilde{\boldsymbol{\vartheta}}}, \boldsymbol{\phi}_{\tilde{\boldsymbol{\vartheta}}} \in \R^D$ and the low-dimensional representation $\mathbf{g} = \mathbf{g}_A , \mathbf{g}_\phi \in \R^K$.
The relation takes the following form:
\begin{align}
	\mathbf{g} = H (\mathbf{f} - \boldsymbol{\mu}) \label{eq:PCA_reduction_model}\\
	\mathbf{f} = H^T \mathbf{g} + \boldsymbol{\mu} \label{eq:PCA_reconstruction_model}
\end{align}
where $\boldsymbol{\mu}$ is the empirical mean vector ${\boldsymbol{\mu} = \frac{1}{N} \sum_{i=1}^N \boldsymbol{f}_i \in \R^D}$ and the matrix $H$ is computed from the empirical covariance ${\Sigma = \frac{1}{N} \sum_{i=1}^N (\mathbf{f}_i-\boldsymbol{\mu}) (\mathbf{f}_i-\boldsymbol{\mu})^T}$.

\subsection{Regression}
Once a dimensional reduction (and reconstruction) scheme is available, we want to perform the regression
\begin{equation} \label{eq:regression_model}
	{\tilde{\boldsymbol{\vartheta}}} \longmapsto \boldsymbol{g}({\tilde{\boldsymbol{\vartheta}}}).
\end{equation}
A number of ML models are available for this purpose. The model Mixture of Experts (MoE) \cite{Jacobs1991AdaptiveMoE} \cite[ch. 11]{murphy2012machine} is found to be a good compromise between simplicity and flexibility.

MoE performs the following 1D regression:
\begin{equation} \label{eq:MoE}
	y(\mathbf{x}) = \sum_{l=1}^L (W^T \mathbf{x})_l \cdot \mathcal{S}(V^T\mathbf{x})_l \ ,
\end{equation}
where $\mathcal{S}$ is the \textit{softmax function}:
\begin{equation} \label{eq:softmax}
	\mathcal{S}(V^T{\mathbf{x}})_l = \frac{e^{(V^T{\mathbf{x}})_l}}{\sum_{l^\prime = 1}^L e^{(V^T{\mathbf{x}})_{l^\prime}}} \ ,
\end{equation}
and ${\mathbf{x}} \in \R^{\tilde{M}}$ and $V,W \in \mathbf{Mat}(\tilde{M},L)$.
The meaning of Eq.~\eqref{eq:MoE} is clear: the output is a weighted combination of $L$ linear regressions $(W^T \mathbf{x})_l$ (called \textit{experts}); each expert performs a reliable regression in a small region of the space. The softmax function (in this context also called \textit{gating function}) switches on the expert contributions whenever this is required.
MoE is usually fitted with the Expectation Maximization (EM) algorithm, which iteratively sets the $W$ and $V$ by refining a lower bound to the log-likelihood of the model.

Linear regression is a very simple model, often inadequate to model a complex relation. A simple trick to improve its performance is called \textit{basis functions expansion}. It consist in the replacement:
\begin{equation}
	{\mathbf{x}} \longrightarrow {\boldsymbol{\xi}}({\mathbf{x}}) = [\xi_1({\mathbf{x}}), \ldots, \xi_M({\mathbf{x}})]^T \ .
\end{equation}
Thus, each expert becomes a non linear regression of the input ${\mathbf{x}}$.
A careful choice of basis functions can really make a difference in fit performances 
and it must be done at validation time, by comparing performances of different models.

The user must choose the number $L$ of experts and the basis functions features ${\boldsymbol{\xi}}({\tilde{\boldsymbol{\vartheta}}}) \in \R^{M}$ to use.
Including in the $\xi_i$ every monomial up to 3rd or 4th order in the three variables $ (\log q, s_1, s_2)$ seems a good working choice for our model
The choice of working with the variable $\log q$ rather than $q$ is based on validation results.
Heuristically, it prevents the values of the data features from varying too much within the 
range of interest, thus yielding more stable numerical performance.

As MoE model deals with single dimensional outputs, a single independent regression must be performed for each component $g_k$ of $\mathbf{g} \in \R^K$
\footnote{This is not a great limitation, because, due to orthogonality of PCs, each $g_j$ is independent from the other: we do not miss correlation among different regressions.}.
In general, a regression will be a collection of MoE weights ${\{ W^{(k)}, V^{(k)} \in \mathbf{Mat}(M,L_k) \}_{k=0}^K}$, where index $k$ labels different regressions for each PC.

\subsection{Summary}
The model has the following explicit form:
\begin{align}
	& \textrm{model}: \mathcal{P} \subset \R^3 \rightarrow \R^K \rightarrow \R^D \nonumber\\
	& {\tilde{\boldsymbol{\vartheta}}}
	\longmapsto  \mathbf{g}({\tilde{\boldsymbol{\vartheta}}} ) = 
		\begin{pmatrix}
		\sum_{l=1}^{L_1} (W^{(1)\;T} \boldsymbol{\xi})_l \cdot \mathcal{S}(V^{(1)\;T}\boldsymbol{\xi})_l \\
		\vdots \\
		\sum_{l=1}^{L_K}  (W^{(K)\;T} \boldsymbol{\xi})_l \cdot \mathcal{S}(V^{(K)\;T}\boldsymbol{\xi})_l
		\end{pmatrix}
	\nonumber \\	
	& \qquad \qquad \qquad \longmapsto \mathbf{f}({\tilde{\boldsymbol{\vartheta}}} ) = H^T \mathbf{g}({\tilde{\boldsymbol{\vartheta}}} ) + \boldsymbol{\mu} \label{eq:model}
\end{align}
where ${\boldsymbol{\xi}}({\tilde{\boldsymbol{\vartheta}}}) \in \R^M $ are the chosen basis function for the regression and $\mathcal{S}(\cdot)_k$ is the \textit{softmax} function Eq.~\eqref{eq:softmax}.
Two relations of the same type must be fitted, one for the amplitude, the other for the phase.

Once weights are set properly, the expression provides an estimation for the waveform $h_{\rm FIT}$ in \eqref{eq:h_std}.
The complete WF $h(t;m_1,m_2, s_1, s_2, d_L, \iota, \varphi_0)$ is computed with Eq.~\eqref{eq:h_parametrization_simple}.
The model can extrapolate outside the range of train orbital parameters, without guarantee of reliable results.

Note that Eq.~\eqref{eq:model} can be used to compute a closed form expression for the gradients 
of the waveform with respect to the orbital parameters. Such calculations are included 
in the released version of \texttt{mlgw}.

\section{Model performance}
\label{sec:performance}
We now discuss some validation tests on our model. We first study how its performance depends on the choice of hyperparameters. Second, we assess the model accuracy and its limitations.
Finally, we measure the speed up provided by our model as compared with training EOB model.
For our tests, we train our model with \texttt{TEOBResumS}. Very similar results are obtained for a model trained on \texttt{SEOBNRv4}.

As it is common, we measure the similarity between two waves by means of the \textit{optimal mismatch}:
\begin{align}
	\bar{\mathcal{F}}[h_1,h_2] & = 1- \frac{\langle h_1, h_2 \rangle}{\sqrt{\langle h_1, h_1 \rangle \langle h_2, h_2 \rangle}}, 	\label{eq:mismatch_def}
\end{align}
where, as usual, we defined the \textit{Wiener product} as:
\begin{equation}
	 {\langle h_1, h_2 \rangle} = 4 \int_0^\infty \d{f} \; \frac{\tilde{h}_1^*(f) \tilde{h}_2(f)}{S_n(f)}.
	\label{ew:Wiener}
\end{equation}
In the equation above, $S_n(f)$ is the detector noise curve, the $\tilde{h}$ denotes the Fourier transform of the 
strain $h$ and the $*$ denotes complex conjugation.
In what follows, we always use a flat noise curve (i.e. constant power spectral density for the detector noise).

\subsection{Validation}
\label{sec:validation}
\begin{figure}[!t]
	\centering
	\includegraphics[width=\linewidth,keepaspectratio]{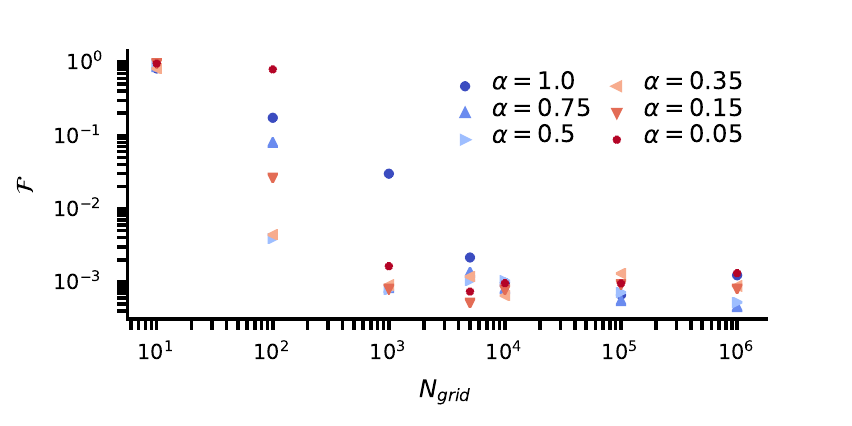}
	\caption{Average mismatch between waves $\mathbf{f}_{N_{\rm grid}, \alpha}$, as saved in the training dataset, and raw waves 
	from EOB model, as a function of time grid size $N_{\rm grid}$. Each series refers to a different values of $\alpha$.
	Clearly, $N_{\rm grid} \simeq 3 \times 10^{3}$ and $\alpha \in (0.3,0.5)$ is a good choice for the dataset hyperparameters.
}
	\label{fig:N_grid}
\end{figure}

Wherever relevant, we will employ a dataset with $5800$ waveforms generated in the domain $\mathcal{P} = [1,20]\times[-0.8,0.95]\times[-0.8,0.95]$, with $\tau_{\rm min} = \SI{1.0}{s/M_\odot}$. The results here refer to \texttt{mlgw-TEOBResumS}, similar results are obtained for \texttt{mlgw-SEOBNRv4}.
\paragraph{Dataset generation parameters}
We first evaluate the impact of number of grid points $N_{\rm grid}$ and distortion parameter $\alpha$ (see Eq.~\eqref{eq:fig01.tau_grid}).
Let $\mathbf{f}_{N_{\rm grid}, \alpha}$ the wave stored in a dataset where $\tau_{\rm min}$ and $\mathcal{P}$ are fixed as above. We compare it with the output of the EOB model $\mathbf{f}_{\rm EOB}$.
We then vary $N_{\rm grid}$ and $\alpha$ and report the resulting mismatch $\mathcal{F}[\mathbf{f}_{\rm EOB}, \mathbf{f}_{N_{\rm grid}, \alpha}]$ in Fig.~\ref{fig:N_grid}.

As expected, we note that, by increasing the number of grid points, the mismatch decreases. Furthermore, using more than $\sim 10^3$ grid points, does not bring any improvement to mismatch. In this case, the result is dominated by numerical errors in the interpolations and it provides a lower-bound for the performances of the fit.
A careful choice of $\alpha$ provides a remarkable improvement when $N_{\rm grid}$ is small. For a high number of grid points, different values of $\alpha$ yield almost equivalent results.
A good setting for dataset hyperparameters might be: $N_{\rm grid} \simeq 3 \times 10^{3}$ and $\alpha \in (0.3,0.5)$.
\begin{figure}[!t]
	\centering
    \begin{minipage}{.5\linewidth}
        \centering
        \includegraphics[width=\linewidth]{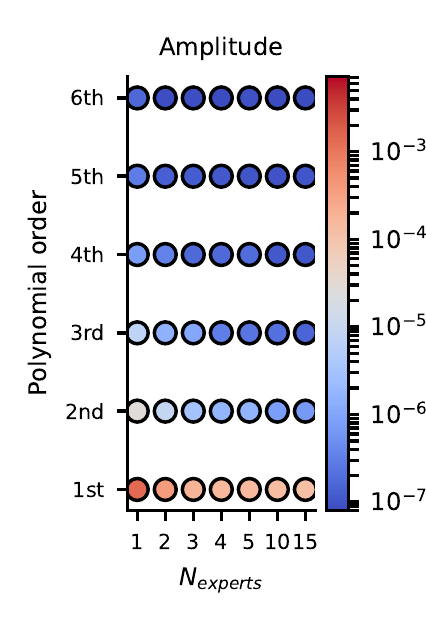}
    \end{minipage}\hfill
    \begin{minipage}{.5\linewidth}
        \centering
        \includegraphics[width=\linewidth]{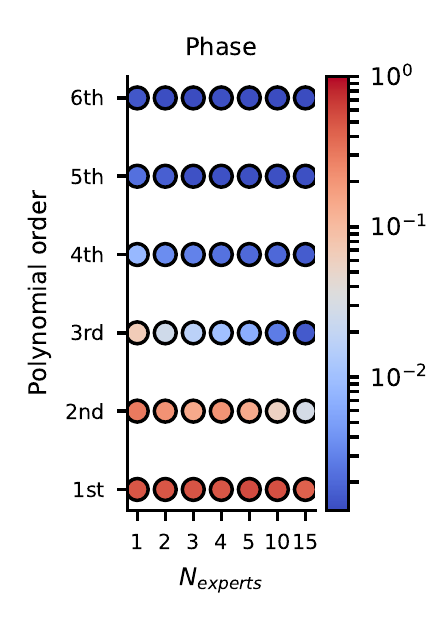}
    \end{minipage}
	\caption{Validation results for fit of MoE model. Each point corresponds to a MoE regressions for the amplitude (left) and phase (right), with a different values of expert number $N_{\rm exp}$ and order of polynomial basis function.
The amplitude and phase are represented with 5 and 4 PCs respectively.
In the colorbar, we represent the mismatch on test waves: it is obtained by reconstructing test waves with fitted amplitude (phase) and test phase (amplitude).
A model with 4 experts and with a 4th order polynomial provides good balance between simplicity and accuracy.
}
	\label{fig:MoE_test}
\end{figure}

\paragraph{MoE parameters}
We only focus on setting the number of experts $N_{\rm exp}$ for each component model and the basis functions $\xi_i(\boldsymbol{\tilde{\vartheta}})$ to use in the regression. Other parameters, related to the details of the training procedure, will not be considered here.

Figure~\ref{fig:MoE_test} presents our results.
We fitted a model for amplitude (or phase) for different configurations of expert number $N_{\rm exp}$ and polynomial basis function.
By label ``n-th order", we mean that in the basis function expansion, every monomial up to $n$-th order is used.
We report with a colorbar the value of the mismatch $F$ between test and reconstructed WFs. The MoE models for each 
component share the same number of experts $N_{\rm exp}$.
The test mismatch for the fitted amplitude (phase) is computed by using the test phase (amplitude) in the reconstructed wave.

As a general trend, fit performance improves whenever the model complexity (i.e. number of fittable parameters) increases.
In general, we note that adding more features is more effective than employing the number of experts.
However, the model performance does not improve indefinitely: as we see in Fig.~\ref{fig:MoE_test}, many ``complex" models show similar performance, regardless their complexity.
A model with 4 experts and 4th order polynomial regression is the ``simplest" of such models and thus it should be deemed as the best choice.
\begin{figure}[!t]
	\centering 
    \begin{minipage}{.5\linewidth}
		\centering
	    \includegraphics[width=\linewidth]{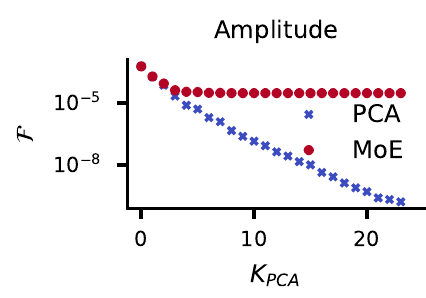}
	\end{minipage}\hfill
    \begin{minipage}{.5\linewidth}
		\centering
	    \includegraphics[width=\linewidth]{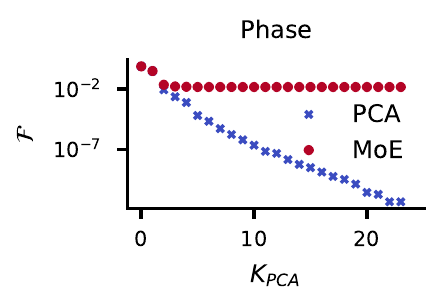}
	\end{minipage}
	\caption{Test mismatch as a function of the number of PCs used in the low dimensional representation.
Label ``PCA" refers to waves reconstructed with PCA only; points with label ``MoE" are reconstructed after a MoE regression.
Data refers to amplitude (left panel) and phase (right panel).
MoE model is chosen to be the optimal one, with $4$ experts and a fourth order polynomial.
}
	\label{fig:mismatch_MoE_vs_PCA}
\end{figure}
\paragraph{Choosing the number of PCs}
Of course, the accuracy of the reconstruction of the low dimensional representation depends on the number $K$ of principal components considered: the more PCs are used, the best accuracy can be achieved.
However in practice, due to errors in the MoE regression, one cannot reduce the reconstruction mismatch
arbitrarily. Indeed, at high PC order the relations to fit become noisy and the regression becomes less accurate, eventually washing out any improvement brought by a higher number of PCs.
For this reason one should choose the number of PCs while checking MoE performance.

In Fig.~\ref{fig:mismatch_MoE_vs_PCA} we report a numerical study of this. We plot the reconstruction mismatch as a function of the number of PCs considered. We consider separately the amplitude and the phase. In one series, we reconstruct the wave using true values of PCs: the mismatch is a measure of PCA accuracy. In the other, we reconstruct a wave using values for PCs as guessed by MoE regression: this is a measure of accuracy of both PCA and regression.
For the first two PCs, the regression is accurate enough for reproducing the PCA accuracy.
On the other hand, any regression beyond the 3rd or 4th PCA component does not give any improvement to the MoE mismatch: the noise in the relation of high order PCs is too high for a regression to be performed.

\begin{figure}
	\centering
    \includegraphics[width=\linewidth]{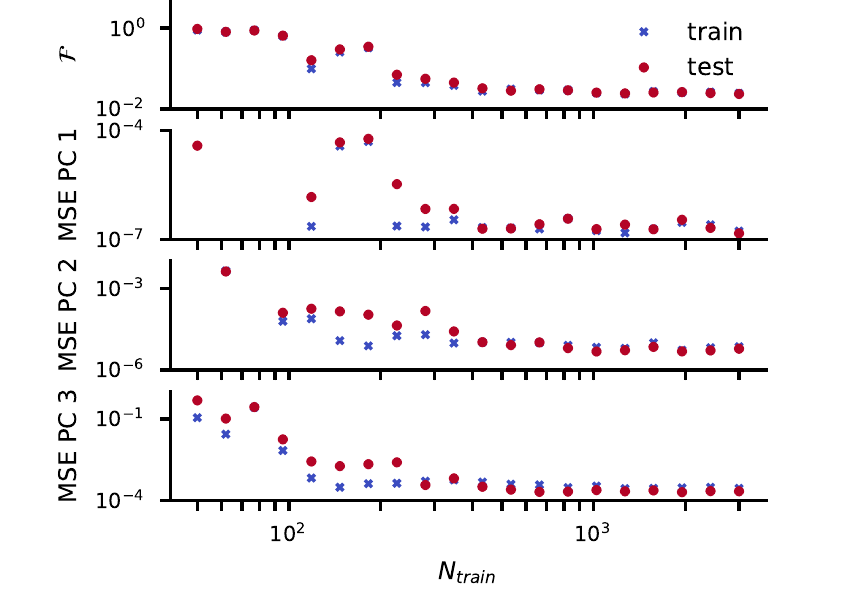}
	\caption{Train and test error for MoE fit of 4 PCs of phase, as a function of the number of training points. 
	We report train and test reconstruction mismatch (top) and mean squared error (MSE) for the first 3 PCs (below). 
	MoE model employs 4 experts and a fourth order polynomial for a basis function expansion. Test mismatch are obtained using 
	test amplitude to reconstruct the waveform; this is not a great limitation as any error in phase reconstruction dominates the overall mismatch.
    }
	\label{fig:N_train}
\end{figure}

In the PCA, we include every PC which yields improvement in MoE mismatch. For our model, $K = 5(4)$ is a good choice for amplitude (phase).
Of course, this strongly depends on the regression model: the more precise the model is, the more PCs can be included.
However, no model can increase its accuracy indefinitely, because every training set has an intrinsic noise level, due to numerical error and to the approximations in the underlying physical model.
\paragraph{Choosing the number of training points}
\label{par:N_train}
The choice of the number of training points $N_{\rm train}$ must trade between accuracy and speed of execution. 
Too many training points will make the training slow, while too few training points will yield a poor model, which does 
not generalize the data ({\it underfitting}). In the choice of number of training points, the comparison between train and 
test error will provide important information on how the model is able to generalize the trend.
In Fig.~\ref{fig:N_train} we report train and test value of mismatch and mean squared error (MSE) of the 
first 3 PCs as a function of the number of training points. Data refers to a MoE model fitted for 4 PCs of the phase dataset, with $4$ experts and a 4th order polynomial. 
As $N_{\rm train}$ increases, we see a steady decrease of the errors, until a plateau is reached.
Since for a reasonably high number of training points ($N_{\rm train} \gtrapprox 50$) train and test error 
are close to each other, we note that overfitting is not a problem.
For $N_{train} \gtrapprox 800$, the trend stabilizes and increasing training points 
does not affect much model performance. In the present model, setting $N_{\rm train} \simeq \SI{3000}{}$ 
is a good choice~\footnote{As compared with standard neural networks, which routinely employ $O(10^5)$ 
points datasets, this is an incredibly low amount of data. This is due to the fact that MoE is a simple 
model with a few number of parameters: few data are enough for learning a reliable relation.}.
\subsection{Accuracy}
\begin{figure}
	\centering
    \includegraphics[width=\linewidth]{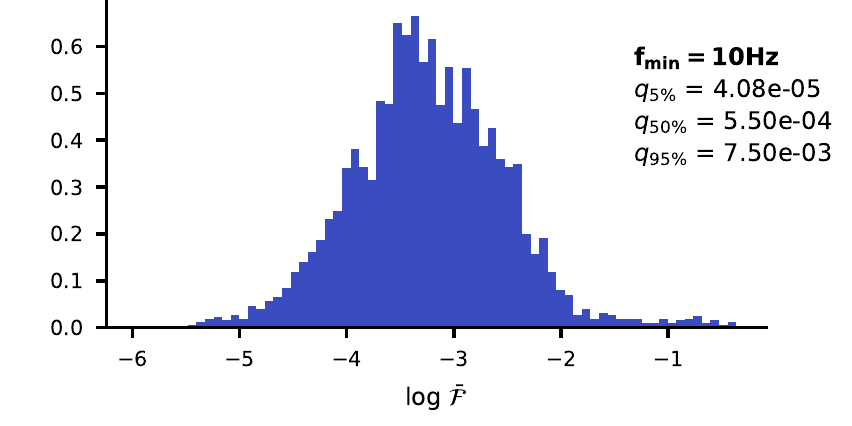}
	\caption{Logarithm of mismatch between \texttt{TEOBResumS} and \texttt{mlgw-TEOBREsumS}, computed on $N=4000$ test waveforms. 
	Each WF is generated with random masses and spins and with a starting frequency of $\SI{10}{Hz}$.
	The median value $q_{50\%}$ and the positions $q_{5\%}$ and $q_{95\%}$ of the 5th and 95th percentile are reported.
}
	\label{fig:F_hist}
\end{figure}
We compute the mismatch between \texttt{mlgw} and the underlying training model (\texttt{TEOBResumS}) for a large number of WFs
and we report our results in the histogram in Fig.~\ref{fig:F_hist}. The mismatch distribution has 
a median mismatch ${\mathcal{F}_m = 5.5 \times {10^{-4}}}$. Such results are similar to the discrepancies 
between state-of-the-art EOB waveforms and NR waveforms~\cite{Bohe:2016gbl, Nagar:2018zoe, Nagar:2020pcj}.

\begin{figure}[t]
	\newcommand{\minipagesize}{.5}
	\newcommand{\titlefontsize}{7}
	\centering 
    \begin{minipage}{\linewidth}
        \centering
	 	\begin{minipage}{\minipagesize\linewidth}
		    \centering
			{\fontfamily{DejaVuSans-TLF} \fontsize{\titlefontsize}{12}\selectfont 
			Mismatch%
			} \\
			\vskip0.3em
		    \includegraphics[width=\linewidth]{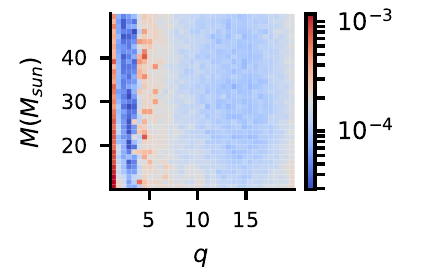}
		\end{minipage}\hfill
		\begin{minipage}{\minipagesize\linewidth}
		    \centering
			{\fontfamily{DejaVuSans-TLF} \fontsize{\titlefontsize}{12}\selectfont 
			Mean Squared Error%
			} \\
			\vskip0.3em
		    \includegraphics[width=\linewidth]{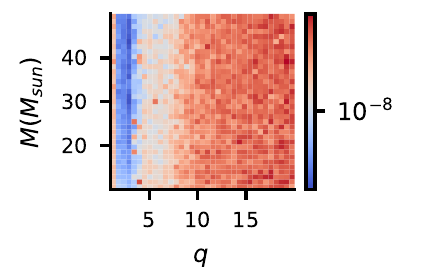}
		\end{minipage}
    \end{minipage}\hfill
    \begin{minipage}{\linewidth}
        \centering
	 	\begin{minipage}{\minipagesize\linewidth}
		    \centering
		    \includegraphics[width=\linewidth]{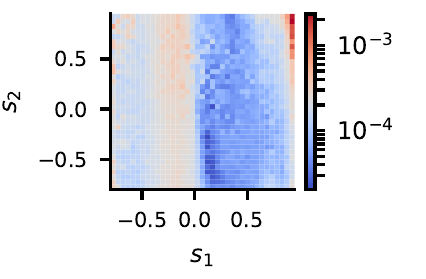}
		\end{minipage}\hfill
		\begin{minipage}{\minipagesize\linewidth}
		    \centering
		    \includegraphics[width=\linewidth]{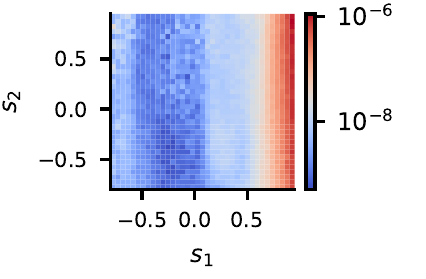}
		\end{minipage}
    \end{minipage}\hfill
    \begin{minipage}{\linewidth}
        \centering
	 	\begin{minipage}{\minipagesize\linewidth}
		    \centering
		    \includegraphics[width=\linewidth]{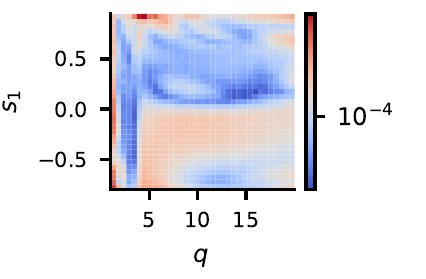}
		\end{minipage}\hfill
		\begin{minipage}{\minipagesize\linewidth}
		    \centering
		    \includegraphics[width=\linewidth]{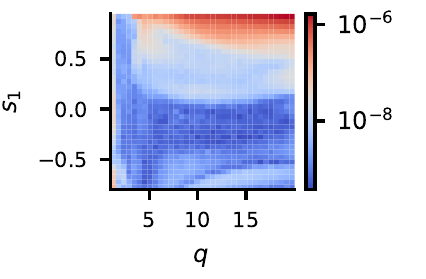}
		\end{minipage}
    \end{minipage}
	\caption{To compare \texttt{TEOBResumS} and \texttt{mlgw-TEOBREsumS}, we report test mismatch (left column) and mean squared errors (MSE) (right column) for the first PC of the phase, as a function of masses and spins. The histograms hold $145061$ waveforms, with randomly drawn parameters.
	Each WF starts $\SI{8}{s}$ before merger.
	Apart from poor performances for ${q\simeq 1}$ and for high (positive) values of $s_1+s_2$, the model performance does not depend much on the input parameters.
}
	\label{fig:countour}
\end{figure}

To understand better model performances, it is interesting to display the accuracy as a function of the 
orbital parameters $\boldsymbol{\vartheta} = (q,M,s_1,s_2)$.
We generate waves for randomly chosen values of $\vartheta = (q, M, s_1, s_2)$ and, for each wave, 
we measure test mismatch $\mathcal{F}$ and MSE on the reconstruction of the first PC for the phase.
The latter is useful to test the accuracy of the fit alone, before wave reconstruction. The results are reported 
in Fig.~\ref{fig:countour}.

The model shows poor performances ($\mathcal{F} \sim 10^{-3}$) for ${q\simeq 1}$ and for high (positive) values of $s_1+s_2$.
By looking at the top line of Fig.~\ref{fig:countour}, we note that the MSE does not depend on $M$,
as expected since the dependence on $M$ is inserted analytically in the model. In the center line of Fig.~\ref{fig:countour}, 
as long as the $s_1$ dependence is considered, the most striking feature is the inverse correlation of mismatch and MSE 
for the first phase PC. This means that, being non-leading, spin contributions are not important for the first PC, 
but become dominant at higher order of PCs. Indeed, the values of the first PC are well correlated with mismatch 
in the case of $q$. See~\cite{Ohme2013PCA_GW} for a closely related discussion on PCA components 
and its dependence on physical parameter.
In the third row of Fig.~\ref{fig:countour}, the inverse correlation between 
the mismatch and MSE can be noted again.
\subsection{Runtime analysis} 
\label{sec:runtime}
We now asses the time performances of our model.
We are interested to make comparisons between
{\tt mlgw} and both training models as well as 
with \texttt{SEOBNRv4\_ROM}.
\paragraph{Comparison with \texttt{TEOBResumS} and \texttt{SEOBNRv4}}
When dealing with a real detection scenario, we are often interested in generating a WF which starts from a given frequency 
$f_{\rm min}$, which is usually set by the detector sensitivity window. Thus, it is crucial to measure the speed up that 
our model can provide in performing such task. We define the speed up $\mathcal{S}$ as the ratio between the runtime 
of the benchmark model and the runtime of \texttt{mlgw} to produce the a waveform starting from a given $f_{\rm min}$. 
Each waveform is produced with constant total mass $M = 100 M_{\odot}$ and random parameters; the WF is sampled 
at $f_{\rm sam} = \SI{2048}{Hz}$. We consider the two cases with $f_{\rm min} = \SI{5}{Hz}$ and $f_{\rm min} = \SI{20}{Hz}$. 
The first choice refers to the hypothetical lower bound for the sensitivity of the Einstein telescope (ET), while the second 
is close to that of Advanced-LIGO/Virgo. In Fig.~\ref{fig:time_performance_hist} we report the histogram of the 
measured speed up values for both \texttt{TEOBResumS} and \texttt{SEOBNRv4}.
  
\newcommand{\factor}{.9}
\begin{figure}
	\centering
	\begin{minipage}{\factor\linewidth}
	    \includegraphics[width=\linewidth]{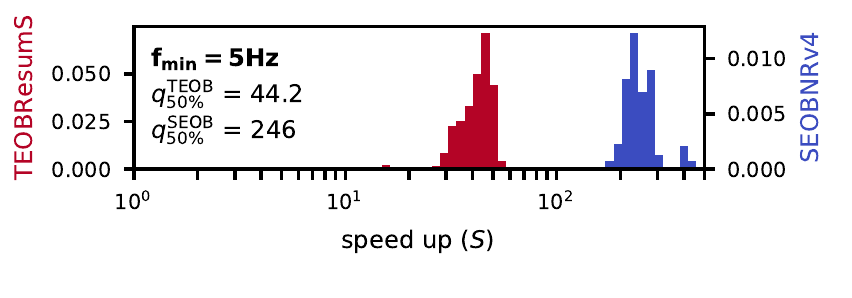}
	\end{minipage}\hfill
	\begin{minipage}{\factor\linewidth}
	    \includegraphics[width=\linewidth]{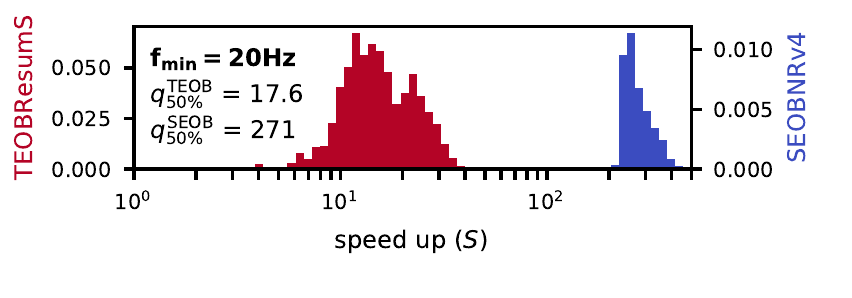}
	\end{minipage}

	\caption{
Speed up given by \texttt{mlgw-TEOBResumS} (blue online) and \texttt{mlgw-SEOBNRv4} (dark-yellow online), 
as compared with their respective native implementation. 
Due to the computational cost, we use $N=500$ test waveforms for \texttt{SEOBNRv4} 
and $N=2000$ waveforms for~\texttt{TEOBResumS}. Each WF is generated with random physical 
parameters and has a minimum frequency of $\SI{5}{Hz}$ (top panel) and $\SI{20}{Hz}$ (bottom panel).
We set a constant total mass $M=\SI{100}{M_\odot}$ and the sampling rate $f_{\rm sam} = \SI{2048}{Hz}$.
Median values $q^{\rm TEOB}_{50\%}$ and $q^{\rm SEOB}_{50\%}$ for the two models are also reported.
}
	\label{fig:time_performance_hist}
\end{figure}
\begin{figure}
	\centering
	\begin{minipage}{\factor\linewidth}
	    \includegraphics[width=\linewidth]{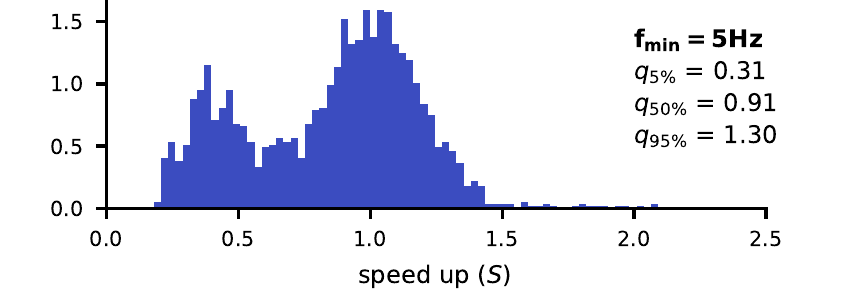}
	\end{minipage}\hfill
	\begin{minipage}{\factor\linewidth}
	    \includegraphics[width=\linewidth]{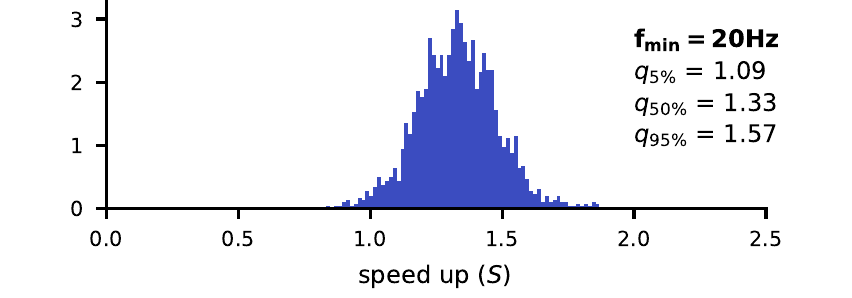}
	\end{minipage}

	\caption{
	Speed up of \texttt{mlgw} with respect to \texttt{SEOBNRv4\_ROM}, computed on $N=2000$ test waveforms.
	Each WF is generated with random physical 
	parameters and has a minimum frequency of $\SI{5}{Hz}$ (top panel) and $\SI{20}{Hz}$ (bottom panel).
	We set a constant total mass $M=\SI{100}{M_\odot}$ and the sampling rate $f_{\rm sam} = \SI{2048}{Hz}$.
	The median value $q_{50\%}$ and the positions $q_{5\%}$ and $q_{95\%}$ of the 5th and 95th percentile are reported.
}
	\label{fig:time_performance_hist_ROM}
\end{figure}

We see that in both cases a substantial speed up is achieved. The speed up is higher for longer WFs, 
making our model particularly convenient 
for advanced detectors, with a larger sensitivity window.
This is clearly understood: a longer WF requires more computation for a EOB model, while 
roughly the same amount of work is done by \texttt{mlgw}.
Furthermore, we note that the speed-up with \texttt{mlgw-SEOBNRv4} is around ten times 
higher than that of \texttt{mlgw-TEOBResumS}. 

\paragraph{Comparison with \texttt{SEOBNRv4\_ROM}}
Let us turn now to discuss a performance comparison with \texttt{SEOBNRv4\_ROM}, that
is currently considered state of the art for the WF generation time.
We note that  \texttt{mlgw-SEOBNRv4}  and \texttt{mlgw-TEOBResumS} are completely
equivalent from the point of view of the generation time for a WF,
so that we simply refer to the model as {\tt mlgw} here and below.
The measured speed up of {\tt mlgw} with respect to \texttt{SEOBNRv4\_ROM} 
is illustrated in Fig.~\ref{fig:time_performance_hist_ROM}.
The comparison is made as above with a sampling rate $f_{\rm sam}=\SI{2048}{Hz}$ for two 
different starting frequency $f_{\rm min} = \SI{5}{Hz}$ and $f_{\rm min} = \SI{20}{Hz}$.
As the ROM model yields WFs in frequency domain, 
in the run-time evaluation we also included a fast Fourier transform (FFT) of the time 
domain WF of \texttt{mlgw}.  This ensures that we are evaluating 
the two model at the same conditions.
Interestingly, the time taken by the FFT (in the \texttt{numpy}
implementation) is similar to that required to generate a WF. Thus for
a WF in FD, our model cannot be substantially faster, due to the
limitation imposed by the FFT~\footnote{Actually, the operations required by
the FFT take the most of the time. In fact,  before the FFT the waveform 
is evaluated on a dense equally spaced grid: as can be seen in 
Table~\ref{tab:profiling}, such operation can be very expensive.}.

We note the the performances are quite similar to each other.
If a lower starting frequency is chosen, \texttt{mlgw} 
is slightly outperformed. Perhaps, this can be cured by fitting a model in frequency domain: 
in this case, the FFT would not be required anymore, resulting in a large speed-up in the execution time.
It is important to stress that \texttt{mlgw}  is written in pure Python, 
while \texttt{SEOBNRv4\_ROM} is coded in $C$.
In fact, a python code could be easily accelerated (i.e. parallelized, run on GPUs, etc...) 
with dedicated libraries, thus allowing to push the code performance further.
%

\paragraph{Profiling}
It is interesting to have a knowledge of the time spent by \texttt{mlgw} 
in each stage of the WF generation procedure. We generate $100$ waves with random 
physical parameters and we measure the CPU time spent to execute each basic task.
In Table~\ref{tab:profiling}, we compare the results for two values of $N_{\rm grid}$.
\begin{table}
	\caption{
Time taken (averaged on multiple runs) by different stages of the generation of $100$ waveforms;
data refers to two different values of $N_{\rm grid}$.
``Generation of raw WF" refers to the computation of the strain $h_{\rm FIT}$ as produced by \texttt{mlgw}. 
``Interpolation to the user grid" evaluates the WF on the grid chosen by the user. 
The ``Post-processing" labels the computation performed to include the dependence on $d_L$, $\iota$ and $\varphi_0$.
}
	\label{tab:profiling}
	\def\arraystretch{1.5}
	\begin{ruledtabular}
	\begin{tabular}{ l c c }
		\multirow{2}{*}{Task (for 100 WFs)}& \multicolumn{2}{c}{CPU time (ms)}\\
			&$N_{\rm grid} = 10^3$	& $N_{\rm grid} = 10^5$\\
	\hline \hline
		 Generation of raw WF 			& $6.9 \; (46.9\%)$	& $7 \; (1.6\%)$ \\ 
	\hline
		 Interpolation to the user grid & $4.5 \; (30.6\%)$	& $194 \; (45.3\%)$ \\ 
	\hline
		 Post processing 				& $1.7 \; (11.6\%)$	& $206 \; (48.1\%)$ \\
	\hline
		 Total							& $14.7 \; (100\%)$ &  $428 \; (100.0\%)$ \\
	\end{tabular}
	\end{ruledtabular}
\end{table}
We see that the cost of generating the raw WF does not depend on the number of grid points. 
On the other hand, the interpolation and the post processing depends on $N_{\rm grid}$ and their 
cost grows dramatically as the user requires more and more points. It is important to stress that 
the latter two tasks are slow only because they deal with a large amount of points. Indeed they perform 
trivial and ``quick" operations and their execution relies on well optimized \texttt{numpy} routines.
If such an amount of datapoints is required, very little space is left for speed up.

\begin{figure*}[t]
	\centering
    \includegraphics[width=\textwidth]{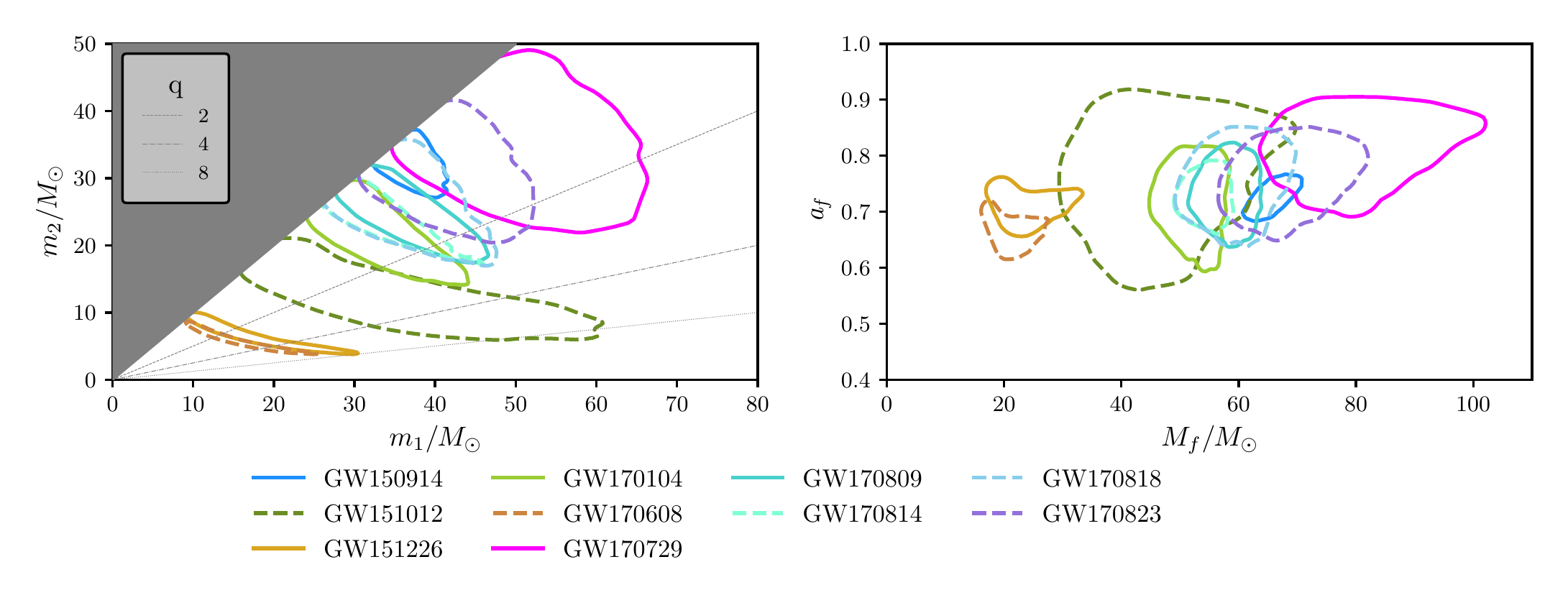}
	\caption{
	Posterior probability densities of the component masses and final masses and spins 
	of all the BBH system in GWTC-1 obtained using MLGW trained with 
	the {\tt TEOBResumS}~\cite{Nagar:2020pcj} spin-aligned waveform model. 
	The contours enclose the $90\%$ credible regions. Left panel: Source-frame component 
	masses $m_1$ and $m_2$. We use the convention $m_1\geq m_2$ which produces 
	the sharp cut in the two-dimensional $(m1,m2)$ distribution (shaded region). Lines of
	constant mass ratio $q\equiv m_1/m_2$ are shown for $q=\{2,4,8\}$. Right panel: the mass $M_f$
	and dimensionless spin magnitude $a_f$ of the final black holes. The figure is consistent with, though 
	different from, Fig.~4 of Ref.~\cite{LIGOScientific:2018mvr}.
}
	\label{fig:gwtc1-summary}
\end{figure*}

\begin{sidewaystable}
\centering
\caption{Summary table for the inferred intrinsic parameters from MLGW with {\tt TEOBResumS} and \texttt{SEOBNRv4} and the released GWTC-1 credible intervals. 
All mass parameters quoted are computed in the source frame, see the text for details of the calculation. For GWTC-1 we report results from Table~III  of Ref.~\cite{LIGOScientific:2018mvr}. These results were obtained by averaging together the outcomes of the 
{\it precessing} {\tt SEOBNRv3}~\cite{Babak:2016tgq} and {\tt IMRPhenomPv2} waveform models. 
The uncertainties correspond to the 90\% credible intervals.
The results with  {\tt TEOBResumS} and  {\tt SEOBNRv4} are very consistent with each other and they are
slightly larger than the published LVC ones obtained using different waveform models.
Note that the inverse mass ratio, $1/q$, is not listed in Ref.~\cite{LIGOScientific:2018mvr}.}
\label{tab:summary}
\begin{ruledtabular}

\begin{tabular}{c|ccccc|ccccc|cccc}
                              & \multicolumn{5}{c}{\texttt{mlgw}-{\tt TEOBResumS}} & \multicolumn{5}{c}{\texttt{mlgw}-{\tt SEOBNRv4}}  & \multicolumn{4}{c|}{GWTC-1}                                                \\ 
                              \hline
Event    & $m_1/M_\odot$ & $m_2/M_\odot$ & $\mathcal{M}/M_\odot$ & $1/q$ & $\chi_{\rm eff}$& $m_1/M_\odot$ & $m_2/M_\odot$ & $\mathcal{M}/M_\odot$ & $1/q$ & $\chi_{\rm eff}$  & $m_1/M_\odot$ & $m_2/M_\odot$ & $\mathcal{M}/M_\odot$ & $\chi_{\rm eff}$
\\ \hline
\vspace{1.0 mm}
GW150914 & $36.36_{-2.64}^{+4.72}$& $32.64_{-4.44}^{+2.93}$& $29.87_{-1.50}^{+1.95}$& $0.91_{-0.21}^{+0.08}$& $0.14_{-0.10}^{+0.10}$&
		 $36.09_{-2.58}^{+4.89}$& $32.55_{-4.37}^{+2.80}$& $29.70_{-1.36}^{+1.95}$& $0.91_{-0.21}^{+0.08}$& $0.10_{-0.08}^{+0.09}$&
    	$35.6_{-3.1}^{+4.7}$  &   $30.6_{-4.4}^{+3.0}$ & $28.6_{-1.5}^{+1.7}$ & $-0.01_{-0.13}^{+0.12}$ \\
\vspace{1.0 mm}
GW151012 & $34.51_{-14.46}^{+21.37}$& $11.67_{-4.46}^{+6.92}$& $16.86_{-2.68}^{+3.01}$& $0.33_{-0.19}^{+0.56}$& $0.53_{-0.33}^{+0.20}$&
		$32.54_{-12.12}^{+20.08}$& $12.18_{-4.50}^{+6.44}$& $16.96_{-2.84}^{+2.67}$& $0.37_{-0.21}^{+0.50}$& $0.53_{-0.32}^{+0.19}$&
		$23.2_{-5.5}^{+14.9}$  &   $13.6_{-4.8}^{+4.1}$ &  $15.2_{-1.2}^{+2.1}$  &  $0.05_{-0.2}^{+0.32}$\\
\vspace{1.0 mm}
GW151226 & $16.44_{-5.52}^{+12.15}$& $6.38_{-2.18}^{+2.86}$& $8.72_{-0.27}^{+0.45}$& $0.39_{-0.24}^{+0.45}$& $0.32_{-0.14}^{+0.24}$& 
		 $16.35_{-5.61}^{+12.60}$& $6.36_{-2.22}^{+2.98}$& $8.69_{-0.27}^{+0.41}$& $0.39_{-0.25}^{+0.48}$& $0.31_{-0.15}^{+0.24}$&
		$13.7_{-3.2}^{+8.8}$& $7.7_{-2.5}^{+2.2}$&  $8.9_{-0.3}^{+0.3}$& $0.18_{-0.12}^{+0.20}$\\
\vspace{1.0 mm}
GW170104 & $31.16_{-4.77}^{+10.55}$& $22.69_{-6.91}^{+4.70}$& $22.83_{-2.06}^{+2.64}$& $0.74_{-0.35}^{+0.23}$& $0.23_{-0.15}^{+0.15}$&
		 $30.45_{-4.56}^{+10.49}$& $22.82_{-7.00}^{+4.43}$& $22.64_{-1.89}^{+2.51}$& $0.76_{-0.37}^{+0.22}$& $0.19_{-0.14}^{+0.15}$&
		$30.8_{-5.6}^{+7.3}$& $20.0_{-4.6}^{+4.9}$& $21.4_{-1.8}^{2.2}$&  $-0.04_{-0.21}^{+0.17}$\\
\vspace{1.0 mm}
GW170608 & $15.45_{-5.05}^{+7.60}$& $5.61_{-1.50}^{+2.32}$& $7.90_{-0.17}^{+0.25}$& $0.36_{-0.18}^{+0.40}$& $0.25_{-0.17}^{+0.20}$&
		 $15.53_{-5.25}^{+8.17}$& $5.58_{-1.57}^{+2.44}$& $7.89_{-0.18}^{+0.25}$& $0.36_{-0.19}^{+0.42}$& $0.24_{-0.18}^{+0.21}$&
		$11.0_{-1.7}^{+5.5}$& $7.6_{-2.2}^{+1.4}$&$7.9_{-0.2}^{+0.2}$&    $0.03_{-0.07}^{+0.19}$           \\
\vspace{1.0 mm}
GW170729 & $50.04_{-9.98}^{+13.97}$& $34.78_{-9.73}^{+9.78}$& $35.73_{-5.08}^{+7.08}$& $0.71_{-0.28}^{+0.26}$& $0.51_{-0.23}^{+0.18}$&
		 $48.51_{-9.62}^{+14.22}$& $34.80_{-9.01}^{+9.26}$& $35.34_{-4.89}^{+6.88}$& $0.73_{-0.27}^{+0.24}$& $0.48_{-0.24}^{+0.20}$&	
		$50.2_{-10.2}^{+16.2}$&  $34.0_{-10.0}^{+9.1}$& $35.4_{-4.8}^{6.5}$&  $0.37_{-0.25}^{+0.21}$\\
\vspace{1.0 mm}
GW170809 & $35.35_{-5.43}^{+8.88}$& $25.17_{-5.91}^{+4.81}$& $25.69_{-1.74}^{+2.35}$& $0.72_{-0.27}^{+0.25}$& $0.24_{-0.14}^{+0.16}$&
		$34.57_{-5.12}^{+7.91}$& $25.23_{-5.47}^{+4.49}$& $25.42_{-1.62}^{+2.41}$& $0.73_{-0.26}^{+0.23}$& $0.20_{-0.13}^{+0.16}$&
		$35.0_{-5.9}^{+8.3}$&$23.8_{-5.2}^{+5.1}$&  $24.9_{-1.7}^{+2.1}$   & $0.08_{-0.17}^{+0.17}$             \\
\vspace{1.0 mm}
GW170814 & $31.35_{-3.48}^{+10.70}$& $25.24_{-6.40}^{+3.16}$& $24.40_{-1.29}^{+1.50}$& $0.81_{-0.36}^{+0.17}$& $0.19_{-0.11}^{+0.12}$&
		$31.53_{-3.73}^{+11.14}$& $24.87_{-6.49}^{+3.29}$& $24.25_{-1.31}^{+1.46}$& $0.79_{-0.36}^{+0.18}$& $0.15_{-0.10}^{+0.11}$&
		$30.6_{-3.0}^{+5.6}$&$25.2_{-4.0}^{+2.8}$&$24.1_{-1.1}^{+1.4}$&   $0.06_{-0.12}^{+0.12}$           \\
\vspace{1.0 mm}
GW170818 &$34.62_{-5.18}^{+11.61}$& $27.09_{-7.80}^{+5.59}$& $26.34_{-2.77}^{+4.03}$& $0.80_{-0.37}^{+0.18}$& $0.28_{-0.19}^{+0.20}$&
		$34.57_{-5.52}^{+11.40}$& $26.94_{-7.67}^{+5.65}$& $26.21_{-2.82}^{+4.32}$& $0.80_{-0.36}^{+0.18}$& $0.25_{-0.17}^{+0.20}$&
		$35.4_{-4.7}^{+7.5}$&$26.7_{-5.2}^{+4.3}$& $26.5_{-1.7}^{+2.1}$&  $-0.09_{-0.21}^{+0.18}$\\
\vspace{1.0 mm}
GW170823 & $40.69_{-6.68}^{+10.21}$& $31.17_{-8.09}^{+7.13}$& $30.63_{-3.75}^{+5.01}$& $0.78_{-0.30}^{+0.19}$& $0.31_{-0.20}^{+0.18}$&
		$40.66_{-6.67}^{+10.11}$& $30.91_{-7.44}^{+6.72}$& $30.52_{-3.59}^{+5.09}$& $0.77_{-0.27}^{+0.20}$& $0.28_{-0.18}^{+0.19}$&
		$39.5_{-6.7}^{+11.2}$&$29.0_{-7.8}^{+6.7}$&   $29.2_{-3.6}^{+4.6}$                    &    $0.09_{-0.26}^{+0.22}$          \\
\end{tabular}

\end{ruledtabular}

\end{sidewaystable}

\section{Application to GWTC-1}
\label{sec:GWTC1}
We use the implementation of our \texttt{mlgw}-{\tt TEOBResumS} and \texttt{mlgw}-{\tt SEOBNRv4} models 
to provide a new, and independent, measure of the properties of the GW sources
collected in GWTC-1, the first catalog of detected GW sources~\cite{LIGOScientific:2018mvr},
corresponding to the first two observing runs of the LIGO and Virgo detectors.
The GWTC-1 catalog consists of 10 BBH systems and a BNS system, GW170817. 
Since the waveform models we considered for the training only concern spin-aligned
BBH waveforms, we do not analyze GW170817 but we only focus on the 10 BBH systems.
We trained \texttt{mlgw} in the range $\mathcal{P} = [1,20]\times[-0.8,0.95]\times[-0.8,0.95]$ 
and we set $\tau_{\rm min} = \SI{4}{s/M_\odot}$.
Our parameter estimation algorithm is \texttt{gwmodel}~\cite{gwmodel} a publicly available infrastructure 
written in a mixture of \texttt{Python} and \texttt{cython} that serves as interface for the parallel nested 
sampling implementation \texttt{cpnest}~\cite{cpnest}. 
The analysis of each BBH system is set up as follows; we choose a total of 2000 Live Points, 
four parallel MCMC chains with a maximum length of 5000 steps to ensure that each successive 
sample is independent of the previous. These settings yield an average of $\sim$ 15000 posterior samples and 
evidence calculations that are accurate to the first decimal digit. For each BBH system we choose 
prior distributions as described in the GWTC-1 release paper~\cite{LIGOScientific:2018mvr}. Finally, 
and critically, to ensure that our results can be compared fairly to published ones, 
we employ the power spectral densities released as part of GWTC-1. No calibration uncertainty model is assumed for these runs. 

\begin{figure*}[t]
	\centering
    \includegraphics[width=\textwidth]{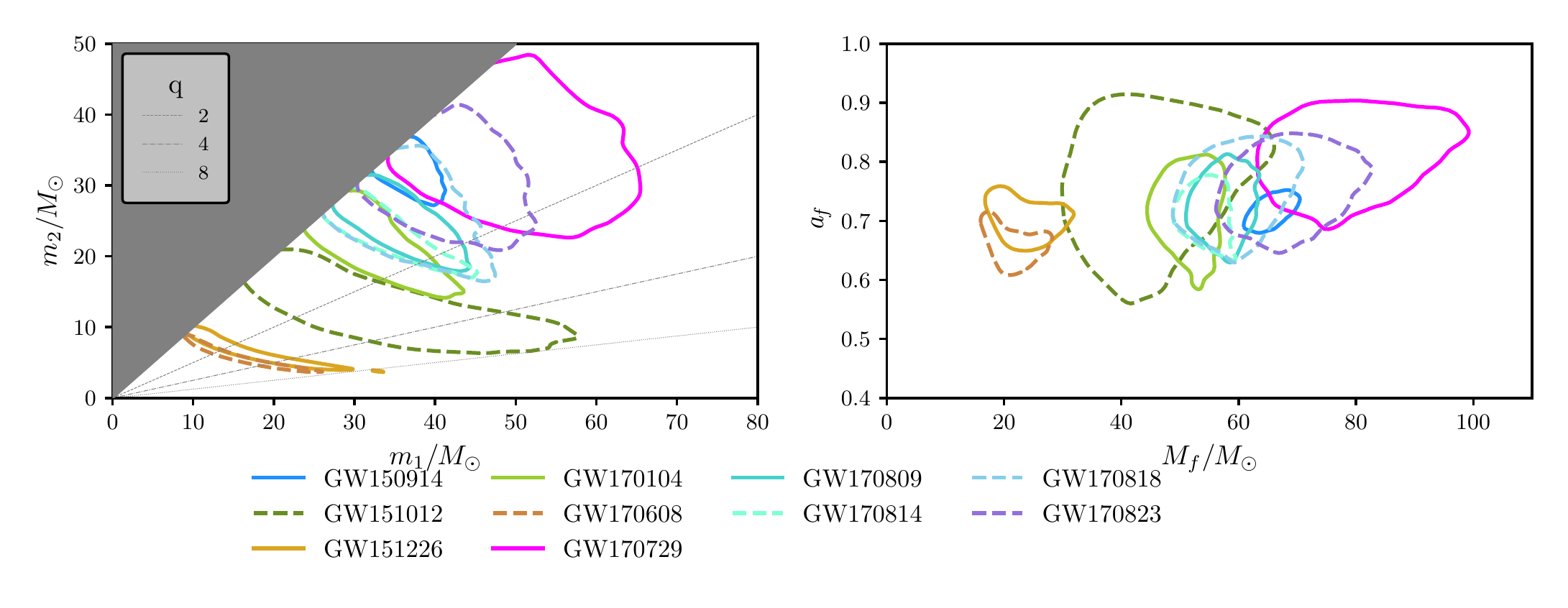}
	\caption{
	Posterior probability densities of the component masses and final masses and spins 
	of all the BBH system in GWTC-1 obtained using MLGW trained with the {\tt SEOBNRv4}~\cite{Bohe:2016gbl}
	waveform model. The contours enclose the $90\%$ credible regions.
	Left panel: Source-frame component masses $m_1$ and $m_2$. We use the convention $m_1\geq m_2$ which 
	produces the sharp cut in the two-dimensional $(m_1,m_2)$ distribution (shaded region). Lines of
	constant mass ratio $q\equiv m_1/m_2$ are shown for $q=\{2,4,8\}$. Right panel: the mass $M_f$
	and dimensionless spin magnitude $a_f$ of the final black holes. The differences with 
	Fig.~\ref{fig:gwtc1-summary} above are practically negligible.}
	\label{fig:gwtc1-summary_seob}
\end{figure*}

\begin{figure*}[t]
	\centering
    \includegraphics[width=\textwidth]{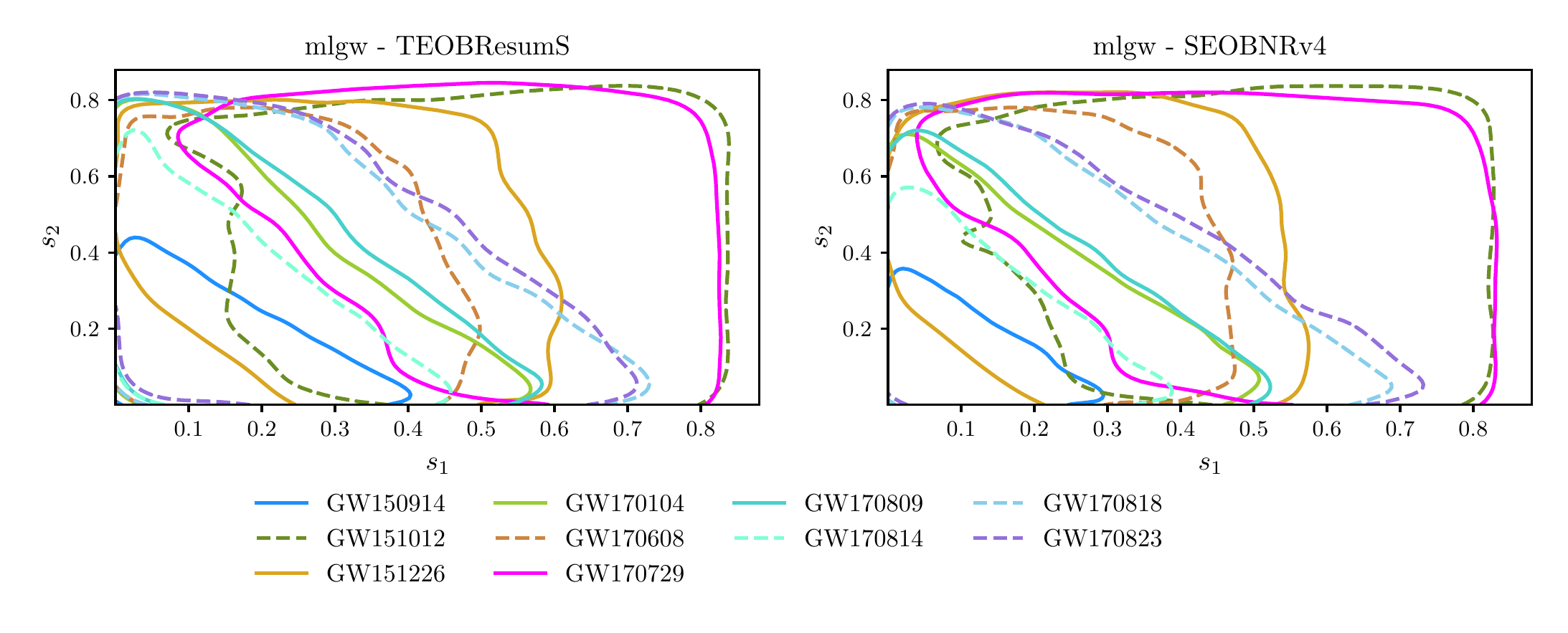}
	\caption{Posterior probability densities of the component, dimensionless, spins for alla the BBH systems in GWTC-1 
	obtained using {\tt mlgw-TEOBResumS}~\cite{Nagar:2020pcj} or {\tt mlgw-SEOBNRv4}~\cite{Bohe:2016gbl}. 
	The contours enclose the $90\%$ credible regions.}
	\label{fig:gwtc1_eob_spins}
\end{figure*}
Table~\ref{tab:summary} summarize the results with {\tt mlgw-TEOBResumS} 
and with {\tt mlgw-SEOBNRv4}. The table exclusively reports summary statistics 
for the intrinsic parameters. All mass parameters quoted are in the source frame. 
The redshift of each BBH is estimated from its luminosity distance posterior and 
converted into a redshift by assuming the cosmological parameters given in Ref.~\cite{Aghanim:2018eyx}.
The second part of Table~\ref{tab:summary} also lists, for convenience, the results
published in the GWTC-1 catalog paper~\cite{LIGOScientific:2018mvr}.
In addition, posteriors for the individual masses, final masses and spins and dimensionless
spin magnitudes are shown in Figs.~\ref{fig:gwtc1-summary}, \ref{fig:gwtc1-summary_seob} and \ref{fig:gwtc1_eob_spins}
for both the models.

A few observation are in order. First of all, our results, obtained with both models,
are extremely similar to what published by the LVK Collaboration. 
This is reassuring as it validates both the WF model hereby presented 
as well as the data analysis scheme and sampler 
implemented\footnote{However, a full validation of the algorithm is presented in ~\cite{gwmodel}.}.
There are however differences that are worth mentioning. The most striking one is that
{\it both} \texttt{mlgw} models tend to recover slightly larger masses and effective 
spin variable, $\chi_{\rm eff}$, than what published Ref.~\cite{LIGOScientific:2018mvr}.
The reason for this discrepancy is probably related to the fact that Ref.~\cite{LIGOScientific:2018mvr}
does not use spin-aligned waveform models, but rather relies the analysis on 
the precessing models \texttt{SEOBNRv3} and \texttt{IMRPhenomPv2}. Although the differences 
are, in general, negligible, still they highlight the differences in the physical input of 
the waveform approximants. By contrast, it is remarkable the excellent agreement between the
two waveform models, although the physical input and the analytical structure of the
two models are rather different, especially in the spin sector~\cite{Rettegno:2019tzh}.
In this respect, we also note in passing that the spin posteriors of GW151012 have
most of the support in the region when $s_1$ and $s_2$ are nonzero. This also
reflects in the rather large value of $\chi_{\rm eff}\sim 0.53$, about one order of magnitude
larger than the result of Ref.~\cite{LIGOScientific:2018mvr}.

\section{Final remarks and future prospects}
\label{sec:end}
We built a ready-to-use Machine Learning model which generates the (dominant quadrupole) time-domain gravitational wave 
signal from a binary Black Hole coalescence in the non precessing case. The code is released as the package \texttt{mlgw}, which is publicly available at 
\href{https://pypi.org/project/mlgw/}{pypi.org/project/mlgw/} and can be installed with the command \texttt{pip install mlgw}.
The model consists of a PCA model to reduce the dimensionality of the $\ell=m=2$ mode (decomposed in amplitude and phase). 
A regression (a MoE model) is performed to infer a relation from the orbital parameters to the low dimensional representation of the WF.

It is important to stress that our model is very simple (i.e. it has a very little number of trainable parameters and 
it is not expensive to train) and flexible (i.e. it works for a large range of parameters and for long waveforms).
In~\cite{Chua_2019}, a ML model for GW generation is built, with similar performances.
However, the reduced dimension space is considerably larger ($O(200)$) than ours ($O(10)$).
In~\cite{Khan:2020fso}, the low dimensional space has a similar dimension $O(30)$ and the ML model achieves a similar performance in the time execution (when computed on a CPU). They manage to achieve a better accuracy $O(2\times10^{-5})$ but they generate significantly shorter WFs ($f_{\rm min} = \SI{15}{Hz}$ for $M = \SI{60}{M_{\odot}}$ against $f_{\rm min} \simeq \SI{2.5}{Hz}$ for \texttt{mlgw}) and need a larger training time ($O(\SI{6}{hours})$ against $O(\SI{6}{minutes})$).

Remarkably, we discovered that a PCA is able to reproduce a high dimensional wave using a small number of variables.
On the other hand, the MoE model is currently the ``bottleneck" of the model accuracy. For this reason, we explored 
several alternative regression methods, including neural networks, but none of them showed dramatically better 
performances: perhaps much more computational power and a larger training set are required to improve any better.

Despite this, our model shows excellent agreement with the underlying training set. At test time, 
the median mismatch is ${\mathcal{F}_m\sim 5 \times 10^{-4}}$.
Furthermore, a single WF generations takes $\SI{0.1}{}-\SI{5}{ms}$ (depending mostly on the 
number of grid points required by the user), which is a factor of $\sim 40$ faster than \texttt{TEOBResumS}
and $\sim 250$ faster than \texttt{SEOBNRv4}.
Interestingly, \texttt{mlgw} matches the performances of a ROM, which is currently close to the state-of-the-art 
for quick generation of waveforms.

The model outputs WFs in time domain. Of course, a similar approach can be applied to WFs in frequency domain: 
this might further speed up the parameter estimation, as the FFT would not be required. A future update to include 
WFs in frequency domain is in program.

Our ML framework allows for several generalization, which might build a more accurate WF generator.
First of all, it is quite straightforward to include higher order modes (HMs) in the WF computation. 
Different regressions, each for each mode, might be done as we already did for the $\ell=m=2$ mode. 
A future update of \texttt{mlgw} along this direction is currently under way.
Second, also the precession effects might be included in the model.
The precession dynamics could be inserted as a single spin parameter $s_P$ \cite{Schmidt2015Precession} 
and the WF dependence on $s_P$ can be fitted together with the other orbital parameters.

Furthermore, our model could be trained on the publicly available NR waveforms 
catalogs (see e.g.~\cite{Mroue:2013xna,Boyle:2019kee,Healy:2019jyf}) and it would provide the best generalization of 
the numerical waveform, dispensing with the EOB models altogether.
Unfortunately, at the moment there are too few NR waveforms ($O(10^2)$) available to perform a reliable 
training: as discussed above in Sec.~\ref{sec:validation}, at least $O(5\times 10^3)$ waveforms are needed:
the improvement shall wait until enough NR waveforms are available.
Moreover, NR waveforms are too short to be used as they are and an extension (e.g. by hybridization with EOB waveforms)
towards the early inspiral is needed to compute any kind of NR-based ML model.

Lastly, we expect our ML approach to work for every kind of source for which a training set of 
waveforms is available. Machine learning models to generate WFs might be crucial in 
the future, where signals from a number of different sources are expected to be detected. In that scenario, 
a parameter estimation must be able to detect among different source and this will require a lot 
of computational work. Speed up will be more pressing.

Our work opens up interesting opportunities in GW data analysis (searches and parameter estimation), 
both because of its speed and of the closed form expression for the WF.

Due to its speed, \texttt{mlgw} could be employed for a systematic comparison between different waveform models, directly on data.
By training (and the training procedure is also quick) \texttt{mlgw} with different waveform models, it will be possible to compare their predictions on several observed events. This could allow to detect systematic biases or to prefer a model over another by means of Bayesian model selection (i.e. by comparing different model evidences).
We started this program by analyzing GWTC-1 with \texttt{mlgw-TEOBResumS} and \texttt{mlgw-SEOBNRv4} and 
highlighted some differences in the predictions as compared to the published results (see Sec.~\ref{sec:GWTC1}). 
Future work might repeat such an analysis on other EOB models or with more observations.

Furthermore, as shown in Fig.~\ref{fig:time_performance_hist}, the model is most useful whenever a long 
waveform is required: in such case, the speed-up gets even more substantial. This is crucial for the detection 
of low frequency signals, as is the case for ET. The analysis of such signals can be performed in the same time 
required to deal with shorter signals: it will become feasible, even with a small amount of resources and 
\textit{without any loss of WF quality}.

A closed form expression for the gradients of the waveform with respect to the orbital parameters
(already included in the \texttt{mlgw} package)
could give an advantage on the parameter estimation procedure by using the Hamiltonian Monte Carlo (HM).
HM \cite{betancourt2017hamiltonianMC}~\cite{Porter2014Hamiltonian_MonteCarlo} is a variant of Markov chain 
Montecarlo, which employs the gradient of the likelihood (dependent on the gradient of the waveform) 
to perform an effective sampling of the posterior distribution. The sampling chain converges faster to 
the steady state by ``finding quickly" the high density regions, thus offering a speed up of the PE.

Another option, so far never explored, is to use the gradients of the WF for a fast exploration of the likelihood landscape. 
With any gradient based optimizer, it should be easy to jump to a \textit{local} maximum of the likelihood. 
Such information might be helpful to reliably locate a \textit{global} maximum of the likelihood.
Such ability could speed up the searches as well as the parameter estimation.

In conclusion, we presented \texttt{mlgw}, an off-the-shelf Machine Learning model for gravitational waves signals from BBHs. 
We demonstrated that \texttt{mlgw} is fast, accurate and easy to train and to use. We anticipate 
that \texttt{mlgw} will enable studies hitherto unfeasible due to the lack of fast and easy to use models.


        \begin{acknowledgments}
         
          R.~G. acknowledges support from the Deutsche Forschungsgemeinschaft
          (DFG) under Grant No. 406116891 within the Research Training Group
          RTG 2522/1. 
          M.~B. and S.~B. acknowledge support by the EU H2020 under ERC Starting
          Grant, no.~BinGraSp-714626.  
          M.~B.~ acknowledges support from the Deutsche Forschungsgemeinschaft
          (DFG) under Grant No. 406116891 within the Research Training Group
          RTG 2522/1. 
          This research has made use of data, software and/or web tools obtained 
          from the Gravitational Wave Open Science Center (https://www.gw-openscience.org), 
          a service of LIGO Laboratory, the LIGO Scientific Collaboration and the 
          Virgo Collaboration. LIGO is funded by the U.S. National Science Foundation. 
          Virgo is funded by the French Centre National de Recherche Scientifique (CNRS), 
          the Italian Istituto Nazionale della Fisica Nucleare (INFN) and the 
          Dutch Nikhef, with contributions by Polish and Hungarian institutes.
        \end{acknowledgments}

	\bibliography{refs20201103.bib}
	\bibliographystyle{ieeetr}

\end{document}